\documentclass[11pt,a4paper]{article}
\usepackage{a4wide}

\usepackage[utf8]{inputenc}
\usepackage[T1]{fontenc}
\usepackage{amsmath}
\usepackage{amssymb}
\usepackage{url}
\usepackage{graphicx}
\usepackage{here}
\usepackage{xparse}
\usepackage{authblk}
\usepackage{amsthm}

\usepackage{adjustbox}

\newtheorem{theorem}{Theorem}
\newtheorem{lemma}[theorem]{Lemma}
\newtheorem{example}{Example}
\newtheorem{observation}{Observation}
\newtheorem{fact}{Fact}
\newtheorem{corollary}{Corollary}

\usepackage{thmtools}
\usepackage{thm-restate}
\newcommand{\Alp}[1][\!]{\Sigma_{#1}}
\newcommand{\minpper}{\mathrm{p}}

\newcommand{\pper}[1][\!]{\parallel_{#1}}

\newcommand{\ps}{\mathsf{PS}}
\newcommand{\psq}{\mathsf{PSeq}}
\newcommand{\Substr}{\mathsf{Substr}}

\newcommand{\floor}[1]{\left\lfloor #1 \right\rfloor}
\newcommand{\ceil}[1]{\left\lceil #1 \right\rceil}

\title{Tight bounds on the number of \\ non-equivalent parameterized squares in a word}
\date{}

\author[1]{Rikuya~Hamai}
\author[1]{Kazushi~Taketsugu}
\author[2]{Yuto~Nakashima}
\author[2]{Shunsuke~Inenaga}
\author[3]{Hideo~Bannai}
\author[4]{Jakub~Radoszewski}

\affil[1]{Department of Information Science and Technology, Kyushu University, Japan}
\affil[ ]{\texttt{hamai.rikuya.226@s.kyushu-u.ac.jp}}
\affil[2]{Department of Informatics, Kyushu University, Japan}
\affil[ ]{\texttt{\{nakashima.yuto.003, inenaga.shunsuke.380\}@m.kyushu-u.ac.jp}}
\affil[3]{M\&D Data Science Center, Institute of Integrated Research, Institute of Science Tokyo, Japan}
\affil[ ]{\texttt{bannai.h.557a@m.isct.ac.jp}}
\affil[4]{Institute of Informatics, University of Warsaw, Poland}
\affil[ ]{\texttt{jrad@mimuw.edu.pl}}

\begin{document}

\maketitle

\begin{abstract}
    Two words $x,y$ of the same length are said to be \emph{parameterized equivalent}
    if there exists a character bijection that transforms $x$ into $y$.
    A word $w$ is called a parameterized square if $w$ is a concatenation of two parameterized equivalent words.    
    Kociumaka et al. [TCS 2016] showed that in a word of length $n$ that contains $\sigma$ distinct characters, the number of \emph{parameterized squares} that are non-equivalent with respect to parameterized equivalence is at most $2 \sigma! n$.
    In this paper, we show that the maximum number of non-equivalent parameterized squares is less than $\sigma n$, which significantly improves the best-known upper bound by Kociumaka et al.
    Moreover, we construct a family of words containing $\Omega(\sigma n)$ non-equivalent parameterized squares,
    which demonstrates that the upper bound is asymptotically tight.
\end{abstract}

\section{Introduction}
In combinatorics on words, repetitive structures (e.g., maximal repetitions, squares) and avoidability are well-studied topics.
Studies on such topics were initiated more than a hundred years ago by Thue~\cite{thue1906uber}.
One of the most fundamental repetitive structures is a square.
A word $w$ is said to be a square if $w$ can be represented as $w = xx$ for some word $x$.
A natural question for squares is revealing the maximum number of squares contained in a word.
Let $\mathit{Sq}(w)$ be the number of distinct squares in a word $w$.
Fraenkel and Simpson~\cite{Fraenkel1998} showed that 
$\mathit{Sq}(w) < 2n$ holds for any word $w$ of length $n$, and conjectured that $\mathit{Sq}(w) < n$,
which subsequently became a long-standing and well-known open question.
This question was positively solved by Brlek and Li~\cite{BrlekL25},
who showed the upper bound $\mathit{Sq}(w) \leq n - |\Alp[w]|$, 
where $\Alp[w]$ denotes the set of distinct characters in $w$.

Kociumaka et al.~\cite{Kociumaka2016} investigated the number of variants of squares under non-standard character-matching models, including parameterized equivalence~\cite{Baker1996}, order-preserving equivalence~\cite{Kim2014,DBLP:journals/ipl/KubicaKRRW13}, and Abelian equivalence~\cite{erdHos1957some}.
Our focus in this paper is the parameterized equivalence model:
For any two length-$k$ words $x$ and $y$ over an alphabet $\Sigma$, $x$ and $y$ are 
said to be parameterized equivalent (denoted by $x \approx y$) if there exists a bijection $f: \Sigma \rightarrow \Sigma$ such that $f(x[i]) = y[i]$ for any position $1 \leq i \leq k$.
For example, $x = \mathtt{ababcdd}$ and $y = \mathtt{cbcbdaa}$ are parameterized equivalent 
because there is a bijection $f$ such that $f(\mathtt{a}) = \mathtt{c}, f(\mathtt{b}) = \mathtt{b}, f(\mathtt{c}) = \mathtt{d}$, $f(\mathtt{d}) = \mathtt{a}$.
Parameterized matching was first introduced by Baker~\cite{Baker1996} 
with motivations to software maintenance,
and various algorithms and data structures were well-studied for pattern matching and other string processing
under the parameterized equivalence model (see~\cite{Amir1994,DeguchiHBIT08,FujisatoNIBT19,0002ST17,Ideguchi2023,IduryS96,IseriIHKYS24,Lewenstein16a,NakashimaFHNYIB22,OsterkampK26} and references therein).
We hereby say that a word $w = xy$ is a \emph{parameterized square} iff $x \approx y$.
For each of these equivalence models, Kociumaka et al.~\cite{Kociumaka2016} introduced two notions of distinctness: one counts non-standard squares that are distinct as words, and the other counts non-standard squares that are non-equivalent under the corresponding equivalence relation.
They showed that any word of length $n$ over an alphabet of size $\sigma$
can contain at most $2 (\sigma!)^2 n$ \emph{distinct} parameterized squares (which are distinct as words),
and at most $2 \sigma! n$ \emph{non-equivalent} parameterized squares~\cite{Kociumaka2016}.

In this paper, we give asymptotically tight upper and lower bounds for the maximum number of non-equivalent parameterized squares in a word.
Let $\psq(w)$ be the number of non-equivalent parameterized squares in a word $w$
(more formal definitions are given in Section~\ref{sec:preliminaries}).

\begin{restatable}{theorem}{Upperbound} \label{thm:equiv-upperbound}
    For any word $w$ of length $n$ that contains $\sigma$ distinct characters, 
    $\psq(w) < \sigma n$ holds. 
\end{restatable}

\begin{corollary} \label{coro:distinct-upperbound}
    For any word $w$ of length $n$ that contains $\sigma$ distinct characters, 
    the maximum number of distinct parameterized squares is less than $\sigma! \cdot \sigma n$.
\end{corollary}

\begin{restatable}{theorem}{Lowerbound} \label{thm:lowerbound}
    There exists an infinite family of words of length $n$ over an alphabet of size $\sigma$ 
    that contains $\Omega(\sigma n)$ non-equivalent parameterized squares.
\end{restatable}

Our result significantly improves the previous upper bound $2 \sigma! n$ by Kociumaka et al.~\cite{Kociumaka2016},
and gives an asymptotically tight lower bound for the upper bound (see Table~\ref{tb:square-variant})
\footnote{Our upper bound was first proposed in the preliminary version~\cite{DBLP:conf/spire/HamaiTNIB24} of this paper,
and the lower bound is a completely new result for this version.}.

\begin{table}[t]
  \centering
  \begin{tabular}{l|c|c|c|c}
    & \multicolumn{2}{c|}{Non-equivalent squares} & \multicolumn{2}{c}{Distinct squares} \\
    Equivalence relation & upper & lower & upper & lower \\ \hline
    Parameterized & & & & \\
    \multicolumn{1}{r|}{(Best known)} & $2\sigma!n$~\cite{Kociumaka2016} & - & $2(\sigma!)^2n$~\cite{Kociumaka2016}  & $\Omega(\sigma n)$ \\
     & & & & (corollary of \cite{GawrychowskiGL23}) \\
    \multicolumn{1}{r|}{(This work)} & $\sigma n$ & $\Omega(\sigma n)$ & $(\sigma!)\sigma n$ & $\Omega(\sigma n)$ \\ \hline
    Order-preserving & $O(\sigma n)$ & - & $O(\sigma n)$~\cite{GawrychowskiGL23} & $\Omega(\sigma n)$~\cite{GawrychowskiGL23} \\
     & (corollary of \cite{GawrychowskiGL23}) & & & \\ \hline
    Abelian & $O(n^{11/6})$~\cite{Kociumaka2016} & $\Omega(n^{1.5})$~\cite{Kociumaka2016} & $O(n^2)$~\cite{Kociumaka2016} & $\Omega(n^2)$~\cite{Kociumaka2016}
  \end{tabular}
  \caption{Three variants of non-standard squares were considered by Kociumaka et al.~\cite{Kociumaka2016}.
  For order-preserving squares, bounds have been improved by Gawrychowski et al.~\cite{GawrychowskiGL23}.
  In this paper, we consider the bounds for non-equivalent parameterized squares.}
  \label{tb:square-variant}
\end{table}
% \begin{table}[t]
%   \centering
%   \begin{tabular}{l|c|c|c|c}
%     & \multicolumn{2}{c|}{Non-equivalent squares} & \multicolumn{2}{c}{Distinct squares} \\
%     Equivalence relation & upper & lower & upper & lower \\ \hline
%     Parameterized equivalence & & & & \\
%     \multicolumn{1}{r|}{(Best known)} & $2\sigma!n$~\cite{Kociumaka2016} & - & $2(\sigma!)^2n$~\cite{Kociumaka2016}  & $\Omega(\sigma n)$~(corollary of \cite{GawrychowskiGL23}) \\
%     \multicolumn{1}{r|}{(This work)} & $\sigma n$ & $\Omega(\sigma n)$ & $(\sigma!)\sigma n$ & $\Omega(\sigma n)$ \\ \hline
%     Order-preserving equivalence & $O(\sigma n)$ (corollary of \cite{GawrychowskiGL23}) & - & $O(\sigma n)$~\cite{GawrychowskiGL23} & $\Omega(\sigma n)$~\cite{GawrychowskiGL23} \\ \hline
%     Abelian equivalence & $O(n^{11/6})$~\cite{Kociumaka2016} & $\Omega(n^{1.5})$~\cite{Kociumaka2016} & $O(n^2)$~\cite{Kociumaka2016} & $\Omega(n^2)$~\cite{Kociumaka2016}
%   \end{tabular}
%   \caption{Three variants of non-standard squares were considered by Kociumaka et al.~\cite{Kociumaka2016}.
%   For order-preserving squares, bounds have been improved by Gawrychowski et al.~\cite{GawrychowskiGL23}.
%   In this paper, we consider the bounds for non-equivalent parameterized squares.}
%   \label{tb:square-variant}
% \end{table}

The \emph{periodicity lemmas}~\cite{Fine1965,Lyndon1962} are the essential tools in the analysis of periodic properties of words under the exact equivalence model.
Apostolico and Giancarlo~\cite{Apostolico2008} presented 
a parameterized version of the periodicity lemma for character bijections which are commutative.
Ideguchi et al.~\cite{Ideguchi2023} proposed a variant of the parameterized periodicity lemma that does not use the commutativity of the bijections.
We also present improved (i.e. tighter) versions of these two parameterized periodicity lemmas, 
which are used for showing our $\sigma n$ bound for the number of non-equivalent parameterized squares in a word.

\subsection{Related work}
Kociumaka et al.~\cite{Kociumaka2016} studied the number of parameterized squares, order-preserving squares, and Abelian squares.
Gawrychowski et al.~\cite{GawrychowskiGL23} gave a tight bound for the number of distinct order-preserving squares.
They also proposed an algorithm to compute all distinct order-preserving squares in a given word of length $n$ in $O(\sigma n)$ time
where $\sigma$ is the alphabet size.
This is optimal time in the worst case thanks to their lower bound result.
Most recently, Nakashima et al.~\cite{DBLP:conf/esa/0001RW25} proposed an algorithm to compute all non-equivalent parameterized squares in a given word in $O(n \sigma \log \sigma)$ time, and an algorithm to count the number of non-equivalent parameterized squares in a given word in $O(n \log n)$ time.

\section{Preliminaries} \label{sec:preliminaries}

\subsection{Words}
Let $\Sigma$ be an {\em alphabet}.
An element of $\Sigma^*$ is called a {\em word}.
The length of a word $w$ is denoted by $|w|$.
The empty word $\varepsilon$ is the word of length 0.
Let $\Sigma^+$ be the set of non-empty words,
i.e., $\Sigma^+ = \Sigma^* \setminus \{\varepsilon \}$.
For any words $x$ and $y$,
let $x \cdot y$ (or sometimes $xy$) denote the concatenation of the two words.
For a word $w = xyz$, $x$ and $y$ are called
a \emph{prefix} and \emph{subword} of $w$, respectively.
Let $\Substr(w)$ denote the set of subwords of $w$.
The $i$-th symbol of a word $w$ is denoted by $w[i]$, where $0 \leq i < |w|$.
For a word $w$ and two integers $0 \leq i \leq j < |w|$,
let $w[i..j]$ denote the subword of $w$ that begins at position $i$ and ends at
position $j$. For convenience, let $w[i..j] = \varepsilon$ when $i > j$.
Also,
let
$w[..i]=w[0..i]$,
$w[i..]=w[i..|w|-1]$, and
$w[i..j] = w(i-1..j] = w[i..j+1)$.
For any word $w$, let $w^1 = w$ and let $w^k = ww^{k-1}$ for any integer $k \ge 2$.
A word $w$ is said to be {\em primitive} if $w$ cannot be written as $x^k$ for any $x \in \Sigma^{+}$ and integer $k \geq 2$.

\subsection{Parameterized equivalence}
Two words $x$ and $y$ of length $k$ each are said to be \emph{parameterized equivalent} 
iff there is a bijection $f$ on $\Sigma$ such that $f(x[i]) = y[i]$ for all $0 \leq i < k$.
We write $x \approx y$ iff two words $x$ and $y$ are parameterized equivalent.
A bijection can be seen as a permutation of $\Sigma$.
A transposition is a permutation that exchanges two elements and leaves every other element fixed.
\begin{fact} \label{fac:permutation}
    Any permutation can be represented by a product of transpositions.
    The parity (odd or even) of the number of transpositions is uniquely determined for any given permutation.
\end{fact}

\subsection{Parameterized squares}
A word $w$ is called a \emph{parameterized square} when $w = xy$ for words $x$, $y$ such that $x$ and $y$ are parameterized equivalent (i.e., $x \approx y$).
We say that two words $x$ and $y$ of equal length are \emph{non-equivalent} (under the parameterized equivalence) iff $x \not \approx y$.
Let $\ps(w)$ denote the set of all parameterized squares occurring in a word $w$.
Let $\psq(w)$ denote the number of equivalence classes on $\ps(w)$ 
with respect to $\approx$.
We call $\psq(w)$ the number of non-equivalent parameterized squares in $w$.

\begin{example}
Let $w = \mathtt{aabbac}$.
For example, $w$ contains
\[
	\mathtt{aa},\mathtt{bb}
\]
as standard squares, 
\[
	\mathtt{aa},\mathtt{ab},\mathtt{ac},\mathtt{ba},\mathtt{bb},\mathtt{aabb},\mathtt{abba}
\]
as parameterized squares, and
\[
	\mathtt{aa} \approx \mathtt{bb}, \mathtt{ab} \approx \mathtt{ac}\approx \mathtt{ba},
 \mathtt{aabb}, \mathtt{abba}
\]
as non-equivalent parameterized squares,
and $\psq(w) = 4$.
\end{example}

\subsection{Parameterized periods}
An integer $p$ is said to be a \emph{parameterized period} (\emph{p-period}) of word $w$
if $w[0..|w|-p) \approx w[p..|w|)$ holds. 
For any p-period $p$ of $w$,
we write $p \pper[f] w$ if $w[0..|w|-p) \approx w[p..|w|)$ holds by a bijection $f$.
We sometimes drop the subscript $f$ (i.e., we just write $p \pper w$) when no confusions occur. 
The smallest p-period of $w$ is denoted by $\minpper(w)$.
By the definition of p-periods, we obtain the following fact which we will use in our proof.

\begin{fact}[cf.~\cite{SCER-period}]
    Let $w$ be a word that satisfies $p \pper[f] w$.
    For any position $i$ satisfying $0 \leq i < |w|-p$, $f(w[i]) = w[i+p]$.
\end{fact}

Apostolico and Giancarlo~\cite{Apostolico2008} showed a variant of the periodicity lemma in the parameterized equivalence model.

\begin{lemma}[Lemma~3 of~\cite{Apostolico2008}] \label{lem:prev-p-perlem1}
    Let $w$ be a word that satisfies $p \pper[f] w$ and $q \pper[g] w$.
    If $p+q \leq |w|$ and $f \circ g = g \circ f$, then $\gcd(p, q) \pper w$.
\end{lemma}

The above lemma uses the commutativity of the two bijections.
Recently, Ideguchi et al.~\cite{Ideguchi2023} proposed a variant of the parameterized periodicity lemma which does not use the commutativity of the bijections as follows:

\begin{lemma}[Lemma~5 of~\cite{Ideguchi2023}] \label{lem:prev-p-perlem2}
    Let $w$ be a word that satisfies $w \in \Sigma^*$, $p \pper w,\, q \pper w$.
    If $p+q +\min(p,q) \cdot (|\Alp[w]|-1) \leq |w|$, then $\gcd(p, q) \pper w$.
\end{lemma}

The number of distinct characters in a word may play an important role in bounding the parameterized periodicity, as can be seen in the previous lemma.
In our result here we will extensively show more of such relations.
The next lemma, shown by Ideguchi et al.~\cite{Ideguchi2023}, also gives a useful relation between them.

\begin{lemma}[Lemma~4 of \cite{Ideguchi2023}]\label{lem:prev-p-per2}
    Let $w$ be a word.
    For any subword $w'$ of $w$, 
    If $|w'| \geq \minpper(w) \cdot (|\Alp[w]|-1)$,
    then $|\Alp[w']| \geq |\Alp[w]|-1$ holds.
\end{lemma}

In the next section, we present a tighter version for each of Lemmas~\ref{lem:prev-p-perlem2} and \ref{lem:prev-p-per2}.

\section{Upper bound for non-equivalent parameterized squares}

We show the following upper bound for the maximum number $\psq(w)$ of non-equivalent parameterized squares in a word $w$.
The case $\sigma=1$ is trivial, since a unary word of length $n$ contains at most $\lfloor n/2 \rfloor$ parameterized-square equivalence classes. Hence, in what follows, we assume $\sigma \ge 2$.

\Upperbound*

To prove Theorem~\ref{thm:equiv-upperbound}, we use a charging argument based on the following theorem:
\begin{restatable}{theorem}{PrefixSquares}\label{lem:prefix-squares}
    For any word $w$ that contains $\sigma$ distinct characters, 
    there can be at most $\sigma$ prefixes of $w$ that are parameterized squares and have no other parameterized occurrence in $w$.
\end{restatable}
For each equivalence class $C$ of parameterized squares occurring in a string $w$ of length $n$, choose an occurrence of a member of $C$ whose starting position is the largest (i.e. rightmost), and charge $C$ to this starting position.
If $C$ is charged to position $i$, then the chosen occurrence is a prefix of the suffix $w[i..]$.
Moreover, by the maximality of $i$, this prefix has no other parameterized occurrence in $w[i..]$;
otherwise, $C$ must have an occurrence starting at a position larger than $i$, a contradiction.
Thus, by Theorem~\ref{lem:prefix-squares}, at most $|\Alp[{w[i..]}]| \leq \sigma$ classes are charged to each position $i$.
Since no parameterized square begins at the last position, summing over all $n-1$ positions gives $\psq(w) \leq \sigma(n-1) < \sigma n$.
It now remains to prove Theorem~\ref{lem:prefix-squares}.

In order to prove Theorem~\ref{lem:prefix-squares}, we first prove Lemma~\ref{lem:substr-character-occ}.
This lemma states that any sufficiently long subword with respect to a p-period of the whole word contains many distinct characters.
Lemma~\ref{lem:substr-character-occ} is a generalized property of a tighter version of the previous statement (Lemma~\ref{lem:prev-p-per2}).

\begin{lemma}\label{lem:substr-character-occ}
    Let $w$ be a word that satisfies $p \pper w$.
    For any subword $w'$ of $w$ and any integer $k$ satisfying $2 \leq k \leq |\Alp[w]| + 1$, 
    if $|w'| \geq p \cdot (k-2) + 1$,
    then $|\Alp[w']| \geq k-1$ holds.
\end{lemma}

\begin{proof}
    We prove this lemma by induction on $k$.
    It is clear that the statement holds for $k=2$.
    Suppose that the statement holds for $k=m$ for some integer $m$ satisfying $2 \leq m \leq |\Alp[w]|$.
    Namely, if $|w'| \geq p \cdot (m-2) + 1$, then $|\Alp[w']| \geq m-1$ holds.
    We show that the statement holds for $k = m+1$.
    Since $|w'| \geq p \cdot (m-1) + 1 \geq p \cdot (m-2) + 1$,
    $|\Alp[w']| \geq m-1$ holds by the induction hypothesis.
    Assume on the contrary that $|\Alp[w']| < m$.
    This implies that $|\Alp[w']| = m-1$.
    Let us consider the prefix $w'[0..i)$ of length $i = p \cdot (m-2) + 1$ of $w'$.
    By the induction hypothesis, $|\Alp[{w'[0..i)}]| \geq m-1$.
    Moreover, since $w'[0..i)$ is a subword of $w'$, $|\Alp[w'{[0..i)}]| = m-1$ holds.
    Due to the p-period $p \pper[f] w$, $f(w'[0..i)) = w'[p..i+p)$ holds.
    This implies that $\Alp[{w'[0..i)}] = \Alp[{w'[p..i+p)}]$ 
    (if not, there exists a character $c \in \Alp[{w'[p..i+p)}] \setminus \Alp[{w'[0..i)}]$
    and then it contradicts $|\Alp[w']| = m-1$).
    Let $A=\Alp[{w'[0..i)}]$.
    Since $p\pper[f] w$ and $\Alp[{w'[p..i+p)}]=A$, we have $f(A)\subseteq A$.
    As $f$ is a bijection and $|f(A)|=|A|$, it follows that $f(A)=A$.
    Consequently, $f^\ell(A)=A$ for every integer $\ell$.
We now prove that every character of $w$ belongs to $A$.
    Since $|w'|\ge p(m-1)+1\ge p+1$, the occurrence of $w'$ in $w$
    contains at least one position from each residue class modulo $p$.
    For any position of $w$, choose a position in this occurrence of $w'$
    with the same residue modulo $p$.
    By repeatedly applying the $p$-period relation forward or backward,
    the corresponding character is obtained by applying some power of $f$
    to a character in $A$.
    Since $f^\ell(A)=A$ for every integer $\ell$, this character also belongs to $A$.
    Thus $\Alp[w]\subseteq A$, which contradicts $|A|=m-1<m\leq |\Alp[w]|$.
\end{proof}

For $|\Alp[w]|\geq 2$, we obtain the following statement by setting $k = |\Alp[w]|$ in Lemma~\ref{lem:substr-character-occ}:
\begin{corollary}\label{coro:substr-character-occ}
    Let $w$ be a word that satisfies $p \pper w$.
    For any subword $w'$ of $w$, 
    if $|w'| \geq p \cdot (|\Alp[w]|-2) + 1$,
    then $|\Alp[w']| \geq |\Alp[w]|-1$ holds.
\end{corollary}

By using this corollary and the next lemma, we can also obtain a tighter version of the periodicity lemma (Lemma~\ref{lem:prev-p-perlem2}) for parameterized words over an alphabet of size more than one.
To prove it, we introduce a relation between the number of distinct characters of a subword and the commutativity of any two permutations of the alphabet.

\begin{lemma}\label{lem:permutation}
    Assume that $|\Sigma| \geq 2$.
    Let $\Sigma'$ be a subset of $\Sigma$ satisfying $|\Sigma'| = |\Sigma|-2$.
    For any permutations $f$ and $g$ of $\Sigma$,
    if $(f \circ g) (a) = (g \circ f) (a)$ for all $a \in \Sigma'$,
    then $f$ and $g$ commute.
\end{lemma}

\begin{proof}
    Let $f$ and $g$ be permutations of $\Sigma = \{c_1, \dots , c_{\sigma}\}$.
    Also, let $\Sigma' = \{c_1, \dots, c_{\sigma-2}\}$ be a subset of $\Sigma$
    such that $(f \circ g) (a) = (g \circ f) (a)$ holds for every $a \in \Sigma'$.
    Assume on the contrary that $(f \circ g) (c_{\sigma-1}) \neq (g \circ f) (c_{\sigma-1})$.
    Let $(f \circ g) (c_{\sigma-1}) = c_i$ and $(g \circ f) (c_{\sigma-1}) = c_j$ for some $i \neq j$.
    Then, $(f \circ g) (c_{\sigma}) = c_j$ and $(g \circ f) (c_{\sigma}) = c_i$.
    Thus, $g \circ f = (c_i, c_j) \circ (f \circ g)$ where $(c_i, c_j)$ is a transposition of elements $c_i$ and $c_j$.
    This implies that $g \circ f$ and $(c_i, c_j) \circ (f \circ g)$ are products of transposition with different parity.
    This fact contradicts Fact~\ref{fac:permutation}.
\end{proof}

\begin{lemma}\label{lem:p-perlem}
    Let $w$ be a word that satisfies $p \pper w$, $q \pper w$, and $|\Alp[w]| \geq 2$.
    If $p + q + \min(p, q) \cdot (|\Alp[w]| - 2) \leq |w|$, $\gcd(p, q) \pper w$ holds.
\end{lemma}
\begin{proof}
    Let $w$ be a word that satisfies $p \pper[f] w$, $q \pper[g] w$, and $|\Alp[w]| \geq 2$.
    Also, let $w' = w[0..\min(p, q) \cdot (|\Alp[w]| - 2))$.
    Then, $(g \circ f)(w') = w[p+q..\min(p, q) \cdot (|\Alp[w]| - 2)+p+q)$ and 
    $(f \circ g)(w')=w[p+q..\min(p, q) \cdot (|\Alp[w]| - 2)+p+q)$.
    Therefore, for any $a \in \Alp[w']$, $(f \circ g) (a) = (g \circ f) (a)$. 
    By Lemma~\ref{lem:substr-character-occ}, $|\Alp[w']| \geq |\Alp[w]|-2$, thus, by Lemma~\ref{lem:permutation}, $f \circ g = g \circ f$.
    Hence, by Lemma~\ref{lem:prev-p-perlem1}, $\gcd(p, q) \pper w$ holds.
\end{proof}

In the proof of Theorem~\ref{lem:prefix-squares}, we need to discuss structures of overlapping parameterized squares.
If there exist two words $xy$ and $yz$ that are overlapping each other and have the same standard period (in the exact matching model) that does not exceed $|y|$, 
we can see that the whole word $xyz$ also has the same period.
This property does not always hold for the parameterized equivalence, 
since p-periods in the two overlapping words may come from different bijections.
In the next lemma, we show that a similar property holds if the overlapping part is sufficiently long (i.e., the overlapping has many distinct characters due to the above lemmas).

\begin{lemma}\label{lem:overlapping-period}
    For any non-empty words $x, y$, and $z$,
    if $p \pper xy$, $p \pper yz$, and $|y| \geq p \cdot (|\Alp[xyz]|-1)+1$,
    then $p \pper xyz$.
\end{lemma}

\begin{proof}
    If $|\Alp[xyz]| = 1$, then $xyz$ is unary and the claim is immediate.
    Hence assume $|\Alp[xyz]| \geq 2$.
    Let $f$ (resp. $g$) be a bijection of $p \pper xy$ (resp. $p \pper yz$), and
    $\sigma = |\Alp[xyz]|$.
    Since $p \pper[f] xy$ and $p \pper[g] yz$ and $|y| \geq p+1$, we have $p \pper[f] y$ and $p \pper[g] y$.
    This implies that $f(y[0..(\sigma-2)p]) = y[p..(\sigma-1)p]$, and
    $g(y[0..(\sigma-2)p]) = y[p..(\sigma-1)p]$.
    Thus, $f(c) = g(c)$ for any character $c \in \Alp[{y[0..(\sigma-2)p]}]$.
    By Lemma~\ref{lem:substr-character-occ}, $|\Alp[{y[0..(\sigma-2)p]}]| \geq \sigma-1$ holds.
    These facts imply that $f = g$.
    Thus, $f((xyz)[0..|xyz|-p)) = (xyz)[p..|xyz|)$ holds and the lemma holds.
\end{proof}

The next lemma explains that a p-period of a prefix which is sufficiently long can extend to the whole word, 
if the whole word has a p-period that is a multiple of the p-period of the prefix.

\begin{lemma}\label{lem:period-extension}
    Let $s$ be a word that satisfies $q \pper[g] s$, and
    $w$ be a prefix of $s$ that satisfies $p \pper[f] w$.
    If $|w| \geq p(|\Sigma_s|-2)+q+1$ and $q = kp$ for some integer $k$, then $p \pper[f] s$.
\end{lemma}

\begin{proof}
    If $|\Alp[s]| = 1$, then $s$ is unary and the claim is immediate.
    Hence assume $|\Alp[s]| \geq 2$.
    Let $w'=w[0..p(|\Sigma_s|-2)]$.
    Since $|w'|+q \leq |w|$ and $q=kp$, 
    $g(w')=w[q..p(|\Sigma_s|-2)+q]=w[kp..p(|\Sigma_s|-2)+kp]=f^k(w')$,
    and hence $g(a)=f^k(a)$ for every $a\in\Alp[w']$.

    Assume on the contrary that $|\Alp[w']| \leq |\Sigma_s|-2$.
    This implies that $|w'| = p(|\Sigma_s|-2)+1 \geq p|\Alp[w']|+1$.
    If $|\Alp[w]| \geq |\Alp[w']|+1$, we can use Lemma~\ref{lem:substr-character-occ} for $w$ and $k = |\Alp[w']|+2~(\leq |\Alp[w]|+1)$.
    Then, we obtain $|\Alp[w']| \geq k-1 = |\Alp[w']|+1$, which is a contradiction.
    Thus, $|\Alp[w]| < |\Alp[w']|+1$ holds.
    By combining with $|\Alp[w']| \leq |\Alp[w]|$, we get $|\Alp[w']| = |\Alp[w]|$.
    Since $g(a)=f^k(a)$ for any $a \in \Alp[w']$, 
    $g(\Alp[w']) = f^k(\Alp[w']) \subseteq \Alp[w] = \Alp[w']$ holds.
    This implies that $g(\Alp[w']) = \Alp[w']$.
    Moreover, we obtain $|\Alp[s]| \leq |\Alp[w']|$ from the fact that $q$ is a p-period of $s$ by the bijection $g$.
    This is a contradiction to the assumption $|\Alp[w']| \leq |\Sigma_s|-2$.
    Thus, $|\Alp[w']| > |\Sigma_s|-2$ holds, and $g=f^k$ over $\Alp[s]$.

    Let $v = w[0..q)$ and $v' = v\left[0..|s|-{\left\lfloor \frac{|s|}{q} \right\rfloor} \cdot q \right)$. 
    Then, 
    \[
        s = v\cdot g(v) \cdot g^2(v) \cdot \cdot \cdot g^{\left\lfloor \frac{|s|}{q} \right\rfloor -1}(v) \cdot g^{\left\lfloor \frac{|s|}{q} \right\rfloor}(v').
    \]

    Let $u=v[0..p)$. Then, $v=uf(u) \cdot \cdot \cdot f^{k-1}(u)$ and 
    \begin{align*}
        s &= v\cdot g(v) \cdot g^2(v) \cdot \cdot \cdot g^{\left\lfloor \frac{|s|}{q} \right\rfloor -1}(v) \cdot g^{\left\lfloor \frac{|s|}{q} \right\rfloor}(v') \\
        &= uf(u) \cdot \cdot \cdot f^{k-1}(u)f^k(u)f^{k+1}(u) \cdot \cdot \cdot f^{2k-1}(u)\cdot \cdot \cdot f^{\left\lfloor \frac{|s|}{p} \right\rfloor}(u'),
    \end{align*}
    where $u'=u\left[0..|s|-{\left\lfloor \frac{|s|}{p} \right\rfloor}\cdot p \right)$.
    Therefore, $p \pper[f] s$.
\end{proof}

\begin{figure}[t]
    \centering
    \includegraphics[keepaspectratio,width=\linewidth]{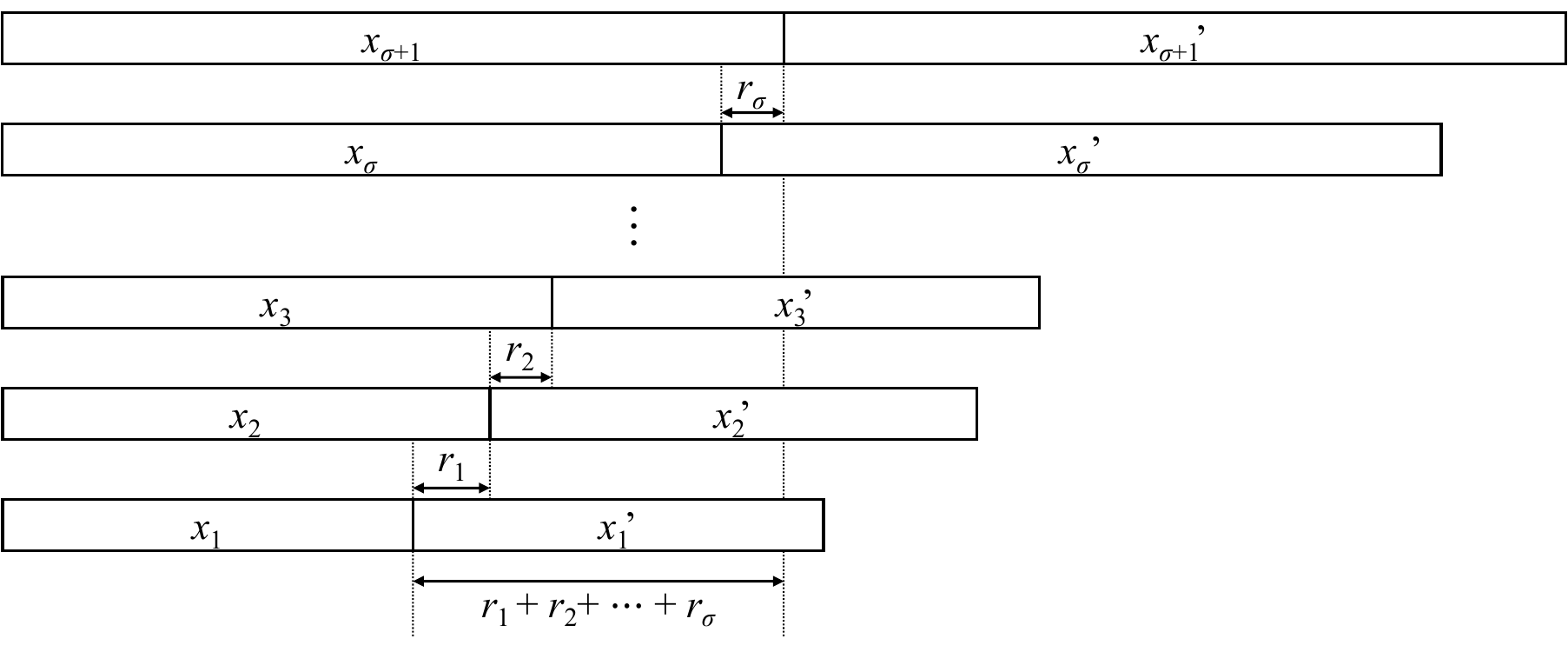}
    \caption{Illustration for the proof of Theorem~\ref{lem:prefix-squares}: $\sigma+1$ non-equivalent parameterized squares cannot begin at the same position.}
    \label{fig:squares}
\end{figure}

Now, we are ready to prove the following main theorem.
\PrefixSquares*

\begin{proof}
    Suppose that there are $\sigma+1$ prefixes of $w$ that are parameterized squares and have no other parameterized occurrence in $w$.
    Let ${x_1}{x_1}', \ldots, {x_{\sigma+1}}{x_{\sigma+1}}'$ 
    denote the $\sigma+1$ parameterized square prefixes that satisfy $|x_1| < \cdots < |x_{\sigma+1}|$
    and $x_i \approx {x_i}'$ for every $i \in [1, \sigma+1]$.
    It is clear from the definition that $|x_{\sigma+1}|/2 < |x_1|$ holds
    (if not, $x_{\sigma+1}'[0..|{x_1}{x_1}'|) \approx {x_1}{x_1}'$).
    We also consider the length $r_k = |x_{k+1}|-|x_k|$ for every integer $k$ satisfying $1 \leq k \leq \sigma$
    (see also Figure~\ref{fig:squares}).
    Since ${x_k}' \approx x_k \approx {x_{k+1}}[0..|x_k|) \approx {x_{k+1}}'[0..|x_k|)$, $r_k \pper x_k$ holds for any $k$.
    By $r_k \pper x_k$, it is clear that $r_k \pper x_1$ for any $k$.
    
    Recall that $\minpper(x_1)$ denotes the smallest p-period of $x_1$. Now we prove $r_\sigma = \ell \cdot \minpper(x_1)$ for some integer $\ell \geq 1$.
    Assume on the contrary that the statement does not hold. 
    From the definitions of $r_1, \ldots, r_{\sigma}$,
    \[
        \minpper(x_1)\cdot(\sigma-1) + r_\sigma \leq r_1 + \cdots +  r_\sigma \leq |x_1|
    \]
    holds.
    This implies that $\minpper(x_1) + r_{\sigma} + \minpper(x_1)\cdot(\sigma-2) \leq |x_1|$.
    Thus, $\gcd(\minpper(x_1), r_\sigma) \pper x_1$ also holds by Lemma~\ref{lem:p-perlem}.
    By the assumption, $\gcd(\minpper(x_1), r_\sigma) < \minpper(x_1)$, which is a contradiction.
    
    In a similar way,
    \[
        \minpper(x_1)\cdot(\sigma-2) + r_\sigma + 1 \leq r_1 + \cdots + r_\sigma \leq |x_1|
    \]
    also holds.
    Hence, this fact and the condition $r_\sigma = \ell \cdot \minpper(x_1)$ for some integer $\ell \geq 1$ imply that $\minpper(x_1) \pper x_\sigma$
    by Lemma~\ref{lem:period-extension}.

    Next, we prove $|x_1| = m \cdot \minpper(x_1)$ for some integer $m \geq 1$.
    Assume on the contrary that the statement does not hold. 
    It is clear that $|x_1| \pper x_{\sigma}$.
    From the definitions of $r_1, \ldots, r_{\sigma}$,
    \[
        \minpper(x_1)\cdot(\sigma-1) + |x_1| \leq |x_1| + r_1 + \cdots + r_{\sigma-1} = |x_{\sigma}|
    \]
    holds.
    This implies that $\minpper(x_1) + |x_1| + \minpper(x_1)\cdot(\sigma-2) \leq |x_{\sigma}|$.
    Thus, by Lemma~\ref{lem:p-perlem}, $\gcd(\minpper(x_1), |x_1|) \pper x_{\sigma}$ also holds.
    By the assumption, $\gcd(\minpper(x_1), |x_1|) < \minpper(x_1)$ holds.
    Moreover, by the definitions of $x_1$ and $x_{\sigma}$,
    $\gcd(\minpper(x_1), |x_1|) \pper x_1$ also holds, which is a contradiction.
    
    It is clear from the definitions that $|x_1| \pper x_{\sigma+1}$.
    In a similar way,
    \[
        \minpper(x_1)\cdot(\sigma-2) + |x_1| + 1 \leq |x_1| + r_1 + \cdots + r_{\sigma-1} = |x_{\sigma}|
    \]
    also holds.
    Hence, this fact and the condition $|x_1| = m \cdot \minpper(x_1)$ for some integer $m \geq 1$ imply that $\minpper(x_1) \pper x_{\sigma+1}$
    by Lemma~\ref{lem:period-extension}.

    We now consider the occurrences of $x_{\sigma+1}$ and ${x_1}'$ in $w$.
    They overlap in an interval of length $r_1+\cdots+r_\sigma$, which is at least $\minpper(x_1)(\sigma-1)+1$.
    Both words have p-period $\minpper(x_1)$: the former by the paragraph above, and the latter because $x_1\approx {x_1}'$ and $\minpper(x_1)\pper x_1$.
    Therefore Lemma~\ref{lem:overlapping-period} yields $\minpper(x_1)\pper x_1{x_1}'$.
    Next consider $x_1{x_1}'$ and ${x_2}'$.
    The overlap length is $|x_1|-r_1$, and since $r_1+\cdots+r_\sigma<|x_1|$ and $r_i\geq\minpper(x_1)$ for $i\geq2$, we have $|x_1|-r_1\geq r_2+\cdots+r_\sigma+1\geq\minpper(x_1)(\sigma-1)+1$.
    Moreover ${x_2}'$ has p-period $\minpper(x_1)$ because $x_2\approx{x_2}'$ and $\minpper(x_1)\pper x_2$.
    Applying Lemma~\ref{lem:overlapping-period} again gives $\minpper(x_1)\pper x_2{x_2}'$.
    Consequently, the substring $x_2{x_2}'[\minpper(x_1)..|x_1{x_1}'|+\minpper(x_1))$ is parameterized equivalent to $x_1{x_1}'$.
    This is another parameterized occurrence of the prefix $x_1{x_1}'$ in $w$, contradicting the assumption.
\end{proof}

Therefore, all proofs are done for our upper bound.

\section{Lower bound for non-equivalent parameterized squares}
In this section, we show a lower bound on non-equivalent parameterized squares.
\Lowerbound*

Let $\alpha \in \mathbb{Z}_+$ and $\lambda = \alpha(\alpha+1)$.
We construct a word $x[0..\lambda)$ over alphabet $\{0,\ldots,\lambda-1\}$ as follows:
\[x[i]=\left.
  \begin{cases}
    \binom{p+1}{2}-1 & \text{if } i=\frac{\lambda}{2}+\binom{p}{2},\ p\in\{1,\ldots,\alpha\}, \\
    i & \text{otherwise}.
  \end{cases}
  \right.\]
Let $\phi$ be a morphism such that $\phi(c)=c^k$ for every $c \in \{0,\ldots,\lambda-1\}$. We define a word $w:=\phi(x)$ of length $\lambda k$. Thus for any $i \in \{0,\ldots,\lambda k - 1\}$, we have
\[w[i]=\left.
  \begin{cases}
    \binom{p+1}{2}-1 & \text{if } \floor{\frac{i}{k}}=\frac{\lambda}{2}+\binom{p}{2},\ p\in\{1,\ldots,\alpha\}, \\
    \floor{\frac{i}{k}} & \text{otherwise}.
  \end{cases}
  \right.\]
We prove this word $w$ gives our lower bound.
We start with simple observations.
\begin{observation}\label{o1}
$\alpha \in \Theta(\sqrt{\lambda})$.
\end{observation}

\begin{observation}
$|\Alp[w]|=|\Alp[x]|=\lambda-\alpha=\alpha^2$.
\end{observation}

\begin{observation}\label{o3}
For every $i \in \{1,\ldots,\lambda-1\}$, we have $x[i] \ne x[i-1]$. Hence, for every $i \in \{1,\ldots,\lambda k - 1\}$, if $i \bmod k = 0$, then $w[i] \ne w[i-1]$.
\end{observation}
\begin{proof}
It suffices to prove the first part of the observation. If $x[i]=i$ and $x[i-1]=i-1$, the conclusion obviously holds. If $x[i] \ne i$, then we consider three cases:
\begin{enumerate}
    \item If $i=\frac{\lambda}{2}$, then $x[i]=0$ but $x[i-1]=\frac{\lambda}{2}-1$.
    \item If $i=\frac{\lambda}{2}+1$, then $x[i]=2$ and $x[i-1]=0$.
    \item If $i > \frac{\lambda}{2}+1$, then $x[i] \le \frac{\lambda}{2}-1$ and $x[i-1]=i-1 > \frac{\lambda}{2}$ since two numbers of the form $\binom{p}{2}$ and $\binom{p+1}{2}$ for positive integer $p$ can differ by 1 only if $p=1$.
\end{enumerate}
In either case, we have $x[i] \ne x[i-1]$.

If $x[i-1] \ne i-1$, the proof is analogous (we only have cases similar to cases 2 and 3 above).
\end{proof}

\begin{figure}[t]
    \centering
    \includegraphics[keepaspectratio,width=\linewidth]{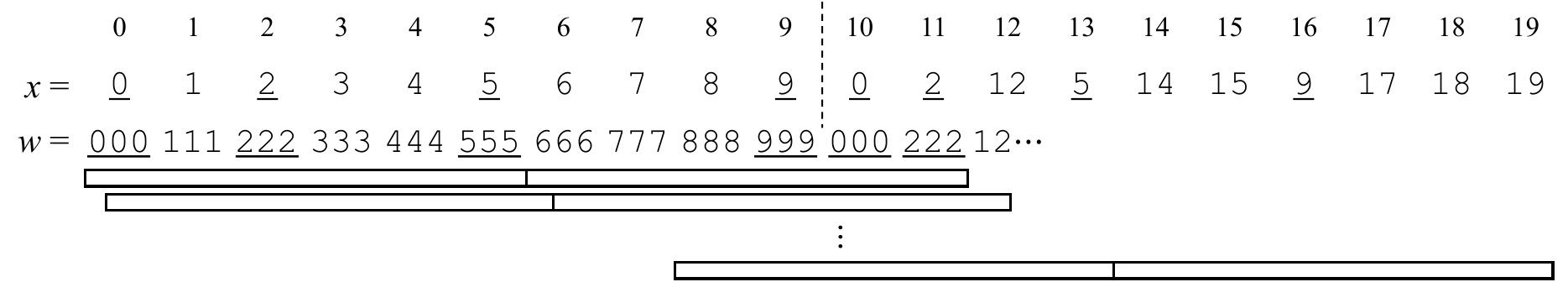}
    \caption{This is an example for case $(\alpha, k) = (4, 3)$. The underlined characters are copied characters (i.e., they have exactly two occurrences in $x$). The word $w$ consists of $k$-copies of each characters. Each pair of boxes indicates a p-square of length $2m = 36$ in $w$.}
    \label{fig:lowerbound}
\end{figure}

The key idea is as follows.
Every length-$2m$ subword whose length is slightly greater than half the length of $w$ is non-equivalent (the precise threshold will be defined in Lemma~\ref{L1}).
This is because, by the definitions of $x$ and $w$, each such subword contains the same character at a special position.
Since the distances between occurrences of this character basically differ for each character, 
all these subwords form parameterized squares for an appropriate $m$.
Figure~\ref{fig:lowerbound} gives an example of a word which we defined and several p-squares in the word.
We show these properties in Lemmas~\ref{L1} and \ref{L2}.

\begin{lemma}\label{L1}
If $2m \ge \frac{\lambda k}{2}+k$, $0 \le i, j \le \lambda k - 2m$, and $i \ne j$, then $w[i..i+2m) \not\approx w[j..j+2m)$.
\end{lemma}
\begin{proof}
We will find indices $a,b \in \{0,\ldots,2m-1\}$ such that
\[(w[i+a]=w[i+b] \,\land\, w[j+a] \ne w[j+b]) \,\lor\, (w[i+a]\ne w[i+b] \,\land\, w[j+a] = w[j+b]).\]
We consider several cases.

\paragraph{Case 1: $i \not\equiv j \pmod{k}$.}
Assume w.l.o.g.\ that $i \bmod k < j \bmod k$. By Observation~\ref{o3}, we have
\[w[j + (k - j \bmod k)] \ne w[j]\text{ but } w[i + (k - j \bmod k)] = w[i]\]
since $\floor{(i + (k - j \bmod k))/k} = \floor{i/k}$.

\paragraph{Case 2: $i \equiv j \pmod{k}$.}
For any index $r$, let
\[p_r = \min\left\{p \in \{1,\ldots,\alpha\}\,:\,k\binom{p+1}{2}-1 \ge r\right\}.\]
Note that $p_i$ and $p_j$ are well-defined as
$i,j < \lambda k-2m < \frac{\lambda k}{2}$.

We have by definition
\begin{equation}\label{eiab}
w\left[k\binom{p_i+1}{2}-1\right]=\binom{p_i+1}{2}-1=w\left[\frac{\lambda k}{2}+k\binom{p_i}{2}+k-1\right].
\end{equation}
Moreover,
\[i \le k\binom{p_i+1}{2}-1 < \frac{\lambda k}{2}\le \frac{\lambda k}{2}+k\binom{p_i}{2}+k-1\le\frac{\lambda k}{2}+i+k-1<i+2m.\]
The penultimate inequality follows by the definition of $p_i$, as $k\binom{p_i}{2}-1 < i$. Hence, \eqref{eiab} can be seen as $w[i+a]=w[i+b]$ for integers $a,b$ such that $0 \le a < b < 2m$ and
\begin{equation}\label{bma}
b-a=\frac{\lambda k}{2}+k\binom{p_i}{2}-k\binom{p_i+1}{2}+k=\frac{\lambda k}{2}-kp_i+k.
\end{equation} 

\paragraph{Subcase 2a: $p_i \ne p_j$.}
Assume w.l.o.g.\ that $p_i < p_j$.
For any integers $a,b$ that satisfy \eqref{bma}, we have $w[j+a] \ne w[j+b]$. Indeed, by definition of $w$, if $w[j+a']=w[j+b']$ for any integers $a',b' \ge 0$, then
\begin{multline*}
b'-a' \le \frac{\lambda k}{2}+k\binom{p_j}{2}+k-1-k\binom{p_j+1}{2}+k=\\ \frac{\lambda k}{2}-kp_j+2k-1\le \frac{\lambda k}{2}-kp_i-k+2k-1 < \frac{\lambda k}{2}-kp_i+k=b-a.
\end{multline*}

\paragraph{Subcase 2b: $p_i = p_j$.}
Assume w.l.o.g.\ that $i<j$. In particular, $j \ge i+k$ as $i \equiv j \pmod{k}$. Then $w[j+a] \ne w[i+a]=\binom{p_i+1}{2}-1$. Similarly as in subcase 2a, if $w[j+a+a']=w[j+a+b']$ for any integers $a',b' \ge 0$, then $b'-a' < b-a$. In particular, $w[j+a] \ne w[j+b]$.
\end{proof}

\begin{lemma}\label{L2}
If $k \mid m$, $i \in \{0,\ldots,\lambda k - 2m\}$, and $m \le k(\frac{\lambda}{2}-\alpha)$, then $w[i..i+2m)$ is a parameterized square.
\end{lemma}
\begin{proof}
Let $a=k-i \bmod k$ and $t=\frac{m}{k}$. We will show that for each $j \in \{i,i+m\}$,
\begin{equation}\label{012t}
w[j..j+m) \approx 0^a 1^k 2^k \dots (t-1)^k t^{k-a}.
\end{equation}
This will imply that $w[i..i+m) \approx w[i+m..i+2m)$ and indeed $w[i..i+2m)$ is a parameterized square.

We proceed to a proof of \eqref{012t}. By the structure of $w$, it is clear that $w[j..j+m) \approx c_0^a c_1^k c_2^k \dots c_{t-1}^k c_t^{k-a}$ for some characters $c_0,c_1,\ldots,c_t$. We only need to show that all of them are different. Assume to the contrary that $c_u=c_v$ for some $0 \le u<v \le t$. By Observation~\ref{o3}, $u+1 \ne v$. This would imply that $w[x]=w[y]$ for some $x<y$ such that $x+k \le y$ and $y-x<m$. However, if $w[x]=w[y]$ for some $x<y$ such that $x+k \le y$, then
\[y-x \ge \min_{p \in \{1,\ldots,\alpha\}} \left\{\frac{\lambda k}{2}+k\binom{p}{2}-k\binom{p+1}{2}+1\right\}=\frac{\lambda k}{2}+1-\alpha k>m.\]
Thus indeed all characters $c_0,c_1,\ldots,c_t$ are different, which implies \eqref{012t}.
\end{proof}

\begin{theorem}\label{t1}
We have $|w|=\alpha(\alpha+1)k$, $|\Alp[w]|=\alpha^2$, and 
the number of non-equivalent parameterized squares in $w$ is at least $\frac{1}{16}\alpha^4 k-O(\alpha^3 k)$.
\end{theorem}
\begin{proof}
By Lemmas~\ref{L1} and~\ref{L2}, the number of non-equivalent parameterized squares in $w$ is at least

\begin{align*}
\psq(w) &\ge \left|\left\{(m,i)\,:\,i \in \left\{0,\ldots,\lambda k - 2m\right\},\, k \mid m,\,\frac{\lambda+2}{4} \le \frac{m}{k} \le \frac{\lambda}{2}-\alpha\right\}\right| \\
&\ge \sum_{t=\ceil{\frac{\lambda+2}{4}}}^{\frac{\lambda}{2}-\alpha} (\lambda k - 2tk + 1) \\
&= \left(\frac{\lambda}{2}-\alpha-\ceil{\frac{\lambda+2}{4}}+1\right)(\lambda k + 1) - 2k \sum_{t=\ceil{\frac{\lambda+2}{4}}}^{\frac{\lambda}{2}-\alpha} t \\
&> \left(\frac{\lambda}{2}-\alpha\right)(\lambda k + 1) - \left(\frac{\lambda+6}{4}\right)(\lambda k + 1)-2k\left(\binom{\frac{\lambda}{2}-\alpha+1}{2}-\binom{\ceil{\frac{\lambda+2}{4}}}{2}\right) \\&= \frac{\lambda^2 k}{4} -2k\left(\frac{\lambda^2}{8}-\frac{\lambda^2}{32}\right) -O(\lambda^{1.5} k) \\
&= \frac{\lambda^2 k}{16}-O(\lambda^{1.5} k).
\end{align*}
Here we used Observation~\ref{o1}.
\end{proof}

To obtain a word over an alphabet of an arbitrary size $\sigma$ that contains $\Theta(\sigma n)$ non-equivalent parameterized squares, where $n$ is the length of the word, it suffices to construct the word $w$ from Theorem~\ref{t1} for $\alpha^2$ being the largest integer square not exceeding $\sigma$ and append $w$ with $\sigma-\alpha^2 \le 2\alpha$ distinct characters.

\section{Conclusions}
We presented asymptotically tight upper and lower bounds for the maximum number of non-equivalent parameterized squares in a word.

An intriguing open question is narrowing the gap between the upper and lower bounds for
the maximum number of parameterized squares which are distinct as words.
As shown in Table~\ref{tb:square-variant},
there is a significant gap between the upper and lower bounds for this problem.

\section*{Acknowledgments}
This work is supported by funding from the Japan Science and Technology Agency (JST) BOOST with grant number JPMJBS2406 (RH), 
the Japan Society for the Promotion of Science (JSPS) KAKENHI with grant numbers
JP25K00136 (YN), JP23K24808, JP23K18466 (SI), and JP24K02899 (HB), and the Polish National Science Center with grant number 2022/46/E/ST6/00463.

\bibliographystyle{abbrv}
\bibliography{ref}

\end{document}